\documentstyle[aps,prd]{revtex}
\newcommand{\f}[2]{\frac{#1}{#2}}
\newcommand{\Ref}[1]{(\ref{#1})}
\newcommand{\be}{\begin{equation}}
\newcommand{\ee}{\end{equation}}
\newcommand{\bn}{\begin{eqnarray}}
\newcommand{\en}{\end{eqnarray}}
\newcommand{\bd}{\begin{displaymath}}
\newcommand{\ed}{\end{displaymath}}
\newcommand{\bnn}{\begin{eqnarray*}}
\newcommand{\enn}{\end{eqnarray*}}
\newcommand{\bml}{\begin{subequations}}
\newcommand{\eml}{\end{subequations}}
\newcommand{\Bx}{\raisebox{.8ex}[1ex][0ex]{\fbox{\rule{.0ex}{0mm}}}}

\begin{document}

\title{ Bremsstrahlung in the gravitational field of a global monopole }
% Short title {Bremsstrahlung in the gravitational field}

\author{Valdir B. Bezerra\thanks{email: valdir@fisica.ufpb.br} and Nail
R. Khusnutdinov\thanks{email: nail@fisica.ufpb.br}\thanks{On leave
from Department of Physics, Kazan State Pedagogical University,
Mezhlauk 1, Kazan 420021, Russia, email: nail@dtp.ksu.ras.ru}}
\address{Departamento de F\'{\i}sica, Universidade Federal da Para\'{\i}ba,\\
Caixa Postal 5008, CEP 58051-970 Jo\~ao Pessoa, Pb, Brasil}
\date{\today}
\maketitle

\begin{abstract}
We investigate the radiation emitted by a uniformly moving charged
scalar particle in the space-time of a point-like global monopole.
We calculate the total energy radiated by the particle and the
corresponding spectrum, for small solid angle deficit. We show
that the radiated energy is proportional to the cube of the
velocity of the particle and to the cube of the Lorenz factor, in
the non-relativistic and ultra-relativistic cases, respectively.
\end{abstract}
%\maketitle

%%%%%%%%%%%%%%%%%%%%%%%%%%%%%%%%%%%%%%%%%%%%%%%%%%%%%%%%%%%%%%%%%%%%%
\section{Introduction}\label{Sec1}
%%%%%%%%%%%%%%%%%%%%%%%%%%%%%%%%%%%%%%%%%%%%%%%%%%%%%%%%%%%%%%%%%%%%%

Topological defects may arise in gauge models with spontaneously
symmetry breaking. They can be of various types as monopoles,
domain walls, strings and their hybrids \cite{Kib76,VilShe94}.
They are cosmological objects which could be formed during phase
transitions in the early universe. They have attracted much
attention because of their peculiar properties, space-time
structure and possible astrophysical implications. Their nature
depends on the topology of the vacuum manifold of the field theory
under consideration.

Among the topological defects mentioned previously, in this paper
we will focus our attention on global monopoles. The simplest
model that gives rise to a global monopole is described by a
system composed of a triplet of isoscalar fields whose original
global {\it O(3)} gauge symmetry was spontaneously broken down to
{\it U(1)}.

The gravitational field of global monopoles may lead to the
clustering in matter and they can induce anisotropies in the
cosmic microwave background radiation. They do not represent any
problem for cosmology and for this reason we may proceed in
studying their physical consequences. On the other hand, in order
to have an agreement with observations, the density of monopoles
has to be very low. The first estimation of the density of
monopoles was made by Hiscock in Ref.\cite{His90} who show that
the upper bound on the number density of the global monopoles is
at most of one monopole in the local group of galaxies. This
estimation was made using the fact that global monopoles produce
enormous tidal acceleration which may be important from the
cosmological point of view. The subsequent numerical simulations
made by Bennet and Rhie show that the upper bound on the density
is smaller than that given by Hiscock by many orders
\cite{BenRhi90}. In fact, one has a scaling solution with a few
global monopoles per horizon volume. This result was recently
recovered by Yamaguchi in Ref.\cite{Yam01}.

Recent observations on the cosmic microwave background anisotropy
in BOOMERANG \cite{Boo} and MAXIMA \cite{Max} experiments present
two peaks in anisotropy, the first is at $l \sim 200$ and the
second one at $l\sim 550$ which are, at the first sight, not
consistent with the locations and width of peaks expected from
topological defects. But as it was noted in
Ref.\cite{BouPetRiaSak00} this statement is partly misleading. The
amplitude of the D\"oppler peaks may be good explained in a hybrid
model, namely, a model which combines inflation with topological
defects. At this moment there is no contradiction with observation
data and there is no observation data which permits rule out
definitively the possibility of existence of topological defects.
Therefore, it is still important from the cosmological as well as
from the astrophysical point of view to investigate different kind
of effects produced in topological backgrounds.

In the framework of Quantum Electrodynamics, the bremsstrahlung process
corresponds to the emission of radiation by a charged particle when it
changes its momentum in collision with obstacles such as other particles
or when it is accelerated due to the presence of electromagnetic fields.
Therefore, in flat space-time particles moving freely do not radiate.
On the other hand, in curved space-time the situation is quite different,
and in this case, a charged particle moving on geodesic does radiate.
This corresponds to the bremsstrahlung process produced by gravitational
fields and this may arises due to the curvature, topology or due to the
combined effects of the geometric and topological features of the space-time.

Alongside with the above radiation emitted by freely moving particles there
is another process leading to radiation. It is well known that
a charged particle placed in a curved space-time, even at the rest,
experiences a self-force due to the geometrical and topological features
of the space-time. In particular, for a conical space-time, it is entirely
due to the non-local structure of the gravitational field\cite{Smi90}.
In the space-time of a point-like global monopole this self-force has already
been calculated \cite{BezFur97} and it is due to the combined
effects of the geometry and topology of this space-time. In this case, the
self-potential has the Coulomb structure and therefore the
particle is repelled by the monopole. Due to this acceleration the
particle emits radiation in a standard way.

In this paper we consider the problem concerning the emission of
radiation by a freely moving particle, caused by the combined
effects of the geometrical and topological features of the
space-time generated by a point-like global monopole. A similar
problem has already been considered in
Refs.\cite{AliGal89,SerSkaFro89,AudJasSka94} in the context of
infinitely thin cosmic strings. It was shown that even freely
moving particles in this space-time emit radiation. The origin of
this radiation is associated with the conical structure of the
cosmic string space-time which produces an effect which is
proportional to the angle deficit. It is worth to mention that
this process is forbidden in the empty Minkowski space-time due to
the energy conservation law.

To begin with, let us first introduce the solution corresponding
to a global monopole considered by Barriola and Vilenkin
\cite{BarVil89}. They considered the simplest model that give rise
to a global monopole which is described by the Lagrangian

\be
L = \f 12 (\partial_\mu \phi^a) (\partial^\mu \phi^a) - \f 14
\lambda (\phi^a \phi^a - \eta^2)^2,
\ee
where $\phi^a$ is a triplet of self-coupling scalar fields and $\eta$ is the
symmetry-breaking scale.

Combining this matter field with the Einstein equations and considering
the general form of the metric with spherical symmetry
\be
ds^2 = B(r) dt^2 - A(r) dr^2 - r^2 (d\theta^2 + \sin^2\theta d\varphi^2),
\ee
the gravitational field is solved and gives the following result

\be
B=A^{-1} = 1 - 8\pi \eta^2 - 2 \f Mr, \label{BA}
\ee
where $M \sim M_{core}$. It is worth noticing that far away from
the global monopole core the main effects are produced by the
solid angle deficit and thus we can neglect the monopole's mass.
Therefore, we obtain the metric of a point-like global monopole
which can be written as

\be
ds^2 = \alpha^2 dt^2 - \alpha^{-2} dr^2 - r^2 (d\theta^2 + \sin^2\theta
d\varphi^2),  \label{MetricGM}
\ee
where the parameter $\alpha$ is connected with the energy scale of
symmetry breaking $\eta$ and is given by the relation $\alpha^2 =
1 - 8\pi \eta^2$. For typical grand unified theory the parameter
$\eta$ is of order $10^{16}GeV$ and thus $1-\alpha^2 = 8\pi \eta^2
\sim 10^{-5}$. The space-time \Ref{MetricGM} is the solution of
Einstein equations with diagonal energy momentum tensor with
components $T^\mu_\nu = diag(2,2,1,1)(\alpha^2 - 1)/r^2$.

Rescaling the time and radial coordinates by relations $t\to t/\alpha^2$ and
$r\to r\alpha^2$ we obtain the following form for the line element
\be
ds^2 = dt^2 - dr^2 - \alpha^2 r^2 (d\theta^2 + \sin^2\theta d\varphi^2)
\label{MetricGM1}
\ee
which will be used in what follows. Here $r\in [0,\infty ],\ \theta\in [0,
\pi],\ \varphi \in [0, 2\pi]$.

This metric corresponds to a space-time with a deficit solid angle
$\Delta=32 \pi^{2} G \eta^{2}$; test particles are deflected
(topological scattering) by an angle $\pi\frac{\Delta}{2}$
irrespective to their velocity and impact parameter. In spite of
having constant coefficients $g_{00} $ and $ g_{rr}$, this metric
represents a curved space-time whose curvature vanishes in the
case $ \alpha=1$ (flat space-time). For $\theta =\frac{\pi}{2}$,
the metric (\ref{MetricGM1}) is exactly the same as that of a
gauge cosmic string, in which case the azimuthal angle $ \varphi$
has a deficit $ \Delta = 2\pi(1- \alpha)$.

Therefore, the gravitational field of a global monopole exhibits
some interesting properties, particularly those concerning the
appearance of nontrivial space-time topologies.

This paper is organized as follows. In Sec. \ref{Sec2} we find an expression
for the spectrum and total energy and analyze these results in two limits:
non-relativistic and ultra-relativistic cases. In Sec \ref{Sec3} we end up
comparing the energy radiated by a freely moving particle and by an
accelerated one due to the self-force and presenting some conclusions.

Throughout this paper we use units $c=G=1$.

%%%%%%%%%%%%%%%%%%%%%%%%%%%%%%%%%%%%%%%%%%%%%%%%%%%%%%%%%%%%%%%%%%%%
\section{The energy and spectrum of radiation.} \label{Sec2}
%%%%%%%%%%%%%%%%%%%%%%%%%%%%%%%%%%%%%%%%%%%%%%%%%%%%%%%%%%%%%%%%%%%%

Now, let us consider a scalar particle with scalar charge $q$
living in this space-time. The scalar and minimal coupling field
corresponding to this particle obeys the Klein - Gordon equation
\be
\Bx \Phi (x) = -4\pi j(x), \label{EqMot}
\ee
with a scalar current
\be
j(x) = q \int \delta^4 (x - x(\tau)) \f{d\tau}{\sqrt{-g}} =
\f{q}{u^0} \f{\delta
(r-r(t)) \delta (\varphi - \varphi (t)) \delta (\theta - \theta (t))}
{\alpha^2
r^2 \sin^2\theta}.
\ee

The trajectory of a freely moving particle in this space-time may be found in
general form from the standard set of equations of geodesic line. For
simplicity and due to spherical symmetry we consider the trajectory of
the particle in the plane $\theta = \f\pi 2$, assuming that  at time $t=0$
the particle is in the closest distance $\rho$ from monopole's core, which
is by definition, the impact parameter. The trajectory has the following
form
\be
r = \sqrt{\rho^2 + v^2 t^2},\ \varphi = \f 1\alpha \arctan \f{vt}\rho ,
\ \theta
= \f\pi 2,\ u^0 \equiv \gamma = \f 1{\sqrt{1 - v^2}},
\ee
where $v$ is the constant velocity of the particle and $u^0$ is the zero
component of the four velocity.

To find the total energy radiated by the particle during all its
history we adopt an approach used in Ref.\cite{AliGal89}. Let us
summarize here its main aspects. The total energy radiated by a
particle is expressed in terms of the covariant divergence of the
energy-momentum tensor as follows
\be
{\cal E} = \int T^\nu_{\mu;\nu} \xi^\mu \sqrt{-g(x)} d^4 x,
\ee
where $\xi^\mu$ is the time-like Killing vector. Taking into account the
explicit form of energy-momentum tensor
\be
T_{\mu\nu} = \f 1{4\pi} \left(\Phi_{,\mu}\Phi_{,\nu} - \f 12 g_{\mu\nu}
\Phi_{,\alpha}\Phi^{,\alpha}\right),
\ee
the equation of motion for a minimally coupled scalar field \Ref{EqMot}
and the explicit expression for the Killing vector $\xi^\mu =(1,0,0,0)$, one
has the following expression for the total energy radiated by the particle
during all time
\be
{\cal E} = 4\pi \int \f\partial {\partial t} D^{rad} (x;x') j(x) j(x')
\sqrt{-
g(x)} \sqrt{-g(x')} d^4 x d^4 x', \label{TotEn}
\ee
where
\be
D^{rad}(x;x') = \f 12 \left[D^{ret}(x;x') - D^{adv}(x;x') \right]
\ee
is the radiative Green function. Here the retarded and advanced Green's
functions $D^{ret}_{adv}(x;x')$ obey the equation
\be
\Bx D^{ret}_{adv}(x;x') = -\delta^4 (x,x').
\ee

In order to find the Green's functions, let us first of all obtain the
complete set of eigenfunctions of the Klein-Gordon equation
\Ref{EqMot}
\be
\Bx \Phi = \left\{\partial^2_t -\f 1{r^2} \partial_r
(r^2\partial_r) + \f 1{\alpha^2 r^2} \hat{L}^2\right\}\Phi =
\lambda^2 \Phi, \label{EqMotMan}
\ee
with eigenvalues $\lambda^2$. Here $\hat{L}^2$ is the square of
the angular momentum operator. The complete set of solution of the
equation \Ref{EqMotMan} was considered in the context of quantum
fields in Ref.\cite{MazLou91}, and it has the following form
\be
\Phi_{l,m,\omega,p} (t,r,\theta,\varphi) = e^{-i \omega t} \sqrt{\f p{2\pi
\alpha^2 r}}
J_{\nu_l} (pr) Y_l^m (\theta,\varphi),
\ee
where $J_\nu(x)$ is the Bessel function of first kind; $Y_l^m(\theta,\varphi)$
is the spherical function $(l=0,1,2,\cdots,\ |m| \leq l))$; $p =
\sqrt{\lambda^2 + \omega^2}$ and
\be
\nu_l = \sqrt{\f{l(l+1)}{\alpha^2} - \f 14}.
\ee

This set of solutions obeys the following relations of orthogonality and
completeness
\bn
\int_0^\infty r^2 dr \int_{-\infty}^{+\infty} dt \int \alpha^2 d\Omega
\Phi_{l,m,\omega,p} (x) \Phi^*_{l',m',\omega',p'} (x) &=&
\delta_{l,l'} \delta_{m,m'} \delta (\omega - \omega') \delta (p-p'),\\
\sum_{l=0}^\infty \sum_{m=-l}^{+l}\int_{-\infty}^{+\infty}
d\omega \int_0^\infty
dp \Phi_{l,m,\omega,p} (x) \Phi^*_{l,m,\omega,p} (x') &=&
\delta (t-t') \f{\delta (r-r')}{\alpha^2 r^2} \f{\delta (\theta - \theta')
\delta (\varphi - \varphi')}{\sin^2 \theta}. \nonumber
\en

Using this set of solutions we may represent the retarded and advanced
solutions in the following form
\be
D^{ret}_{adv}(x;x') = \sum_{l=0}^\infty \sum_{m=-l}^{+l}
\int_{-\infty}^{+\infty} d\omega \int_0^\infty dp
\f{\Phi_{l,m,\omega,p} (x) \Phi^*_{l,m,\omega,p} (x')}{p^2 -
\omega^2 \mp i 0}
\ee
and therefore, the radiative Green function reads
\bn
D^{rad}(x ; x') &=& \f i{2\alpha^2} \f 1{\sqrt{r
r'}}\sum_{l=0}^\infty \sum_{m= - l}^{+l} Y_l^m(\theta,\varphi)
Y_l^{m*}(\theta',\varphi')\int_{-\infty} ^{+\infty}
d\omega\  {\rm sgn} (\omega) e^{-i\omega (t-t')}\\
&\times&\int_0^\infty dp p J_{\nu_l}(pr) J_{\nu_l}(pr')
\delta (p^2 - \omega^2).  \nonumber
\en

Taking into account this formula into Eq. \Ref{TotEn}, we obtain
the following expression for the total energy
\be
{\cal E} = \f{2\pi q^2}{\gamma^2 \alpha^2} \sum_{l=0}^\infty
\sum_{m=-l}^{+l} \left|Y_l^m(\f\pi 2, 0)\right|^2\int_{-
\infty}^{+\infty} d\omega\ |\omega | \int_0^\infty dp p \delta
(p^2 - \omega^2) \left|S_l^m(\omega,p,v,\rho)\right|^2,
\ee
where we have introduced the function $S_l^m$ by the relation
\be
S_l^m(\omega,p,v,\rho) = \int_{-\infty}^{+\infty} d t e^{i\omega t - i \f
m\alpha \arctan \f{vt}\rho} \f{J_{\nu_l}(p\sqrt{\rho^2 + v^2 t^2})}{(\rho^2 +
v^2 t^2)^{1/4}}. \label{DefS}
\ee

This function obeys the following symmetry relation
\be
S_l^m(-\omega,p,v,\rho) = S_l^{-m}(\omega,p,v,\rho).
\ee

Using this we may represent the total energy as an integral
\be
{\cal E} = \int_0^\infty d\omega \f{d {\cal E}}{d\omega}, \label{Intdw}
\ee
where the spectral density is
\be
\f{d {\cal E}}{d\omega} =\omega \f{2\pi q^2}{\gamma^2 \alpha^2}
\sum_{l=0}^\infty \sum_{m=-l}^{+l} \left|Y_l^m(\f\pi 2, 0)\right|^2
\left|S_l^m(\omega,\omega,v,\rho)\right|^2.
\ee

The function $S_l^m(\omega,\omega,v,\rho)$ given by Eq. \Ref{DefS}
may be represented in a slightly different form, more suitable for
analysis (here we assume $\omega >0$) as
\be
S_l^m(\omega,\omega,v,\rho) =
-2 \f{\sqrt{\rho}}v \sin\f\pi 2 \left[\nu_l - \f
m\alpha - \f 12\right] \tilde{S}_l^m(\omega,v,\rho),
\ee
where
\be
\tilde{S}_l^m(\omega,v,\rho) = \int_1^\infty d y e^{-\f {\omega \rho}v y}
\left(\f{y-1}{y+1}\right)^{-\f m{2\alpha}} \f{I_{\nu_l}
(\omega\rho \sqrt{y^2 -
1})}{(y^2-1)^{1/4}}. \label{TilS}
\ee

Therefore we can express the spectral density of radiation by
\be
\f{d {\cal E}}{d\omega} =\omega \rho \f{8\pi q^2}{v^2 \gamma^2 \alpha^2}
\sum_{l=0}^\infty \sum_{m=-l}^{+l} \left|Y_l^m(\f\pi 2, 0)\right|^2
\left|\tilde{S}_l^m(\omega,v,\rho)\right|^2\sin^2\f\pi 2 \left[\nu_l - \f
m\alpha - \f 12\right]. \label{dE/dw}
\ee

Integrating over the frequency $\omega$, using formula 6.612(3) from
Ref.\cite{GraRyz80}, we find that the total energy is
\bn
{\cal E} &=& - \f{8 q^2}{v^3 \gamma^2 \alpha^2 \rho} \sum_{l=0}
^\infty \sum_{m=-
l}^{+l} \left|Y_l^m(\f\pi 2, 0)\right|^2 \sin^2\f\pi 2 \left[\nu_l -
\f m\alpha
- \f 12\right] \label{TotEn1}\\
&\times&\int_1^\infty\f {dy}{y^2 -1}\left(\f{y-1}{y+1}\right)^{-\f
m{2\alpha}}\int_1^\infty\f {dy'}{y'^2 -1}\left(\f{y'-1}{y'+1}\right)^{-\f
m{2\alpha}} (y+y') Q'_{\nu_l - \f 12} [\cosh\sigma]. \nonumber
\en
Here $Q_\nu[x]$ is the Legendre function of second kind; the prime means the
derivative with respect to its argument, and
\bd
\cosh \sigma = \f{(y+y')^2 v^{-2} -y^2 -y'^2 +2}{2\sqrt{y^2-1}\sqrt{y'^2-1}}.
\ed

Now, let us analyze the above expressions in the Minkowski limit.
In this case we have to put $\alpha = 1$. Due to this, the
argument of sine is $\f\pi 2(l- m)$. Next we have to take into
account that $Y_l^m(\f\pi 2, 0) = 0$, if $(l+m)$ is odd. For this
reason the argument of sine is $\f\pi 2\times$ (even number) which
implies that the sine of this quantity is zero and as a
consequence the total energy is zero, too, as it must be in
Minkowski space-time. Differently from the cosmic string
space-time there is no specific values of $\alpha$ for which total
energy is identically zero.

Let us simplify our formulas for the global monopole space-time assuming that
solid angle deficit is small.  In this case we can expand sine in the
previous formulas in terms of $\alpha$ as
\be
\sin^2\f\pi 2 \left[\nu_l - \f m\alpha - \f 12\right] \approx (1-\alpha)^2
\f{\pi^2}4 \left[\f{l(l+1)}{l+ \f 12} -m\right]^2.   \label{SinApp}
\ee

Therefore up to $(1-\alpha)^2$ we may set $\alpha = 1$ in the rest
part. Firstly, let us analyze the total energy given by Eq.
\Ref{TotEn1}. The sum over $m$ can be made using the addition
theorem for Legendre function of the first kind from
Ref.\cite{GraRyz80} and results in
\bn
{\cal E} &=& - \f{\pi q^2 (1-\alpha)^2}{v^3 \gamma^2 \rho} \int_1^\infty\f
{dy}{y^2 -1} \int_1^\infty\f {dy'}{y'^2 -1} (y+y') \sum_{l=0}^\infty
\left[\left(l+\f 12\right)^3 - \f 12 \left(l+\f 12\right) + \f 1{16} \f 1{l+\f
12}\right. \\
&+&\left. 2\left(\left(l+\f 12\right)^2 - \f 14\right) \partial_\beta +
\left(l+\f 12\right)\partial^2_\beta\right]P_l[\cosh\beta] Q'_l [\cosh\sigma],
\nonumber
\en
where
\be
\cosh\beta =\f{yy' +1}{\sqrt{y^2-1}\sqrt{y'^2-1}}.
\ee

Using now an integral representation for the Legendre function of the second
kind as below
\be
Q_l[\cosh \sigma] = \f 1{\sqrt{2}} \int_\sigma^\infty \f{e^{-(l+1/2)
t}dt}{\sqrt{\cosh t - \cosh \sigma}},
\ee
and the relation
\be
\sum_{l=0}^\infty e^{-(l+1/2)t} P_l[\cosh\beta] = \f 1{\sqrt{2}} \f
1{\sqrt{\cosh t - \cosh \beta}},
\ee
we get, finally, the following expression for the total energy
\be
{\cal E} = -(1-\alpha)^2\frac{\pi q^2 v \gamma^2}{2\rho}
\int_1^\infty\int_1^\infty \f{dy dy'}{(y+y')^3} E(s,y,y'),
\ee
where
\bn
E &=& 1 - 48 s^2 z_1 - 192 s^4 z_2^2 -
16 s^2 z_2 \int_0^\infty \f{dx}{\sqrt{x}}
\f\partial {\partial x} \left[\f 1{\sqrt{R_x}}\left\{\f 1{(x+1)^{3/2}} +
\f{6s^2
z_1}{(x+1)^{5/2}} +
\f{15}2\f{s^4 z_2^2}{(x+1)^{7/2}}\right\}\right]\nonumber \\
&+& \f 14 \int_0^\infty \f{dx}{\sqrt{x}} \f\partial{\partial x} \left[ \f
1{\sqrt{R_x}}\int_x^\infty \f{dx'}{\sqrt{R_{x'}}}\f 1{\sqrt{x'+1}}\right],
\label{E}
\en
and we have introduced the following definitions
\bnn
z_1 &=& \f{yy' +1}{(y+y')^2},\ z_2 = \f 1{y+y'},\\
R_x &=& (x+1)^2 + 4s^2 z_1 (x+1) + 4s^4 z_2^2,
\enn
with the parameter $s$ given by $s= v\gamma$.

The formula obtained looks very awesome but it may be analyzed without great
problem for non-relativistic and ultra-relativistic particles. The function
$E$ depends only on the combination $s=v\gamma = v/\sqrt{1-v^2}$. In the non-
relativistic case the parameter $s\to 0$ and in the ultra-relativistic
case $s\to\infty$, and therefore we have to analyze the function $E$ in
these two limits.

In the non-relativistic case we may expand all integrands in Eq.
\Ref{E} over small values of the parameter $s$ and calculate
the integrals. The main contribution, in this case, is
proportional to the cube of the velocity
\be
{\cal E} = (1-\alpha)^2\f{\pi q^2}{2\rho} v^3. \label{ENon}
\ee

The ultra-relativistic case is more complicate due to the last
term in Eq. \Ref{E}. There is no need, in fact, to calculate the
contribution from it. It is enough to find an upper bound for it.
Let us analyze the contribution from this term which is given by
\be
W = -(1-\alpha)^2 \f{\pi q^2 v \gamma^2}{8\rho}\int_1^\infty\int_1^\infty
\f{dy dy'}{(y+y')^3} \int_0^\infty \f{dx}{\sqrt{x}} \f\partial{\partial x}
\left[ \f
1{\sqrt{R_x}}\int_x^\infty \f{dx'}{\sqrt{R_{x'}}}\f 1{\sqrt{x'+1}}\right].
\ee
First of all one represents the polynomial $R$ in the following form
\be
R_x = (x+1 + s^2\delta_+^2)(x+1+s^2\delta_-^2),
\ee
where $\delta_\pm^2 = 2(z_1 \pm \sqrt{z_1^2 - z_2^2})$. Because $\delta_\pm^2$
are positive we can write out the following inequalities
\be
\int_x^\infty \f{dx'}{\sqrt{R_{x'}}} \f 1{\sqrt{x'+1}} \leq \int_0^\infty
\f{dx'}{\sqrt{R_{x'}}} \f 1{\sqrt{x'+1}}
\leq \int_0^\infty \f{dx'}{(x+1)^{3/2}}
\leq 2.
\ee

Using this upper bound we have
\be
|W| \leq (1-\alpha)^2 \f{\pi q^2 v \gamma^2}{4\rho} \int_1^\infty\int_1^\infty
\f{dy dy'}{(y+y')^3} \f{{\bf E}(\sqrt{1-\f{b^2}{a^2}})}{ab^2}
\leq (1-\alpha)^2
\f{\pi^2 q^2v\gamma^2}{8\rho}\int_1^\infty\int_1^\infty \f{dy dy'}{(y+y')^3}
\f{1}{ab^2},
\ee
where $a=\sqrt{1+s^2\delta_+^2},\ b=\sqrt{1+s^2\delta_-^2}$ and we used
the fact
that the upper bound for elliptic integral of second kind ${\bf E}$ is
$\pi/2$. Now we change the variables $y\to s y,\ y'\to s y'$ and take the
ultra-relativistic limit $s\to \infty$. In the end we have the following
estimation
\be
|W| \leq (1-\alpha)^2\f{\pi^3q^2}{32\rho}.
\ee

The contribution of others terms in Eq. \Ref{E} is of order larger
than $\gamma^3$. Calculating the other integrals in Eq. \Ref{E} one has
that the total energy radiated in ultra-relativistic case is given by
\be
{\cal E} =(1-\alpha)^2 \f{3\pi^3q^2}{32\rho} \gamma^3.
\label{EUltra}
\ee

Let us consider the spectral density of energy given by Eq.
\Ref{dE/dw} in the case of small solid angle deficit where
$|1-\alpha| \ll 1$. First of all we change the variable of
integration in Eq. \Ref{Intdw} according to $\omega \to \Omega:\
\Omega = \omega \rho/v$ and use the small solid angle deficit
approximation for sine given by Eq. \Ref{SinApp}. Thus, we obtain
\be
\f{d{\cal E}}{d\Omega} =
\Omega (1-\alpha)^2 \f{2\pi^3 q^2}{ \rho \gamma^2}
\sum_{l=0}^\infty \sum_{m=-l}^{+l} \left|Y_l^m(\f\pi 2, 0)\right|^2 \left[\f
{l(l+1)}{l+1/2} -m \right]^2 \left|\tilde{S}_l^m\right|^2.  \label{SpeDen}
\ee

Now, let us analyze this spectral density in the non-relativistic
limit, that is, $v \ll 1$. To do this, we shift the variable of
integration in Eq. \Ref{TilS}: $x\to x+1$ and use the series
expansion for Bessel function in integrand. Thus, we obtain a
power series over the velocity. Indeed,
\be
\tilde{S}^l_m  = e^{-\Omega} \sum_{k=0}^\infty \f{(\f{\Omega
v}{2})^{l+k+1/2}}{k! \Gamma (l+k+3/2)} \int_0^\infty dx e^{-\Omega x} x^{\f{l-
m}2 + \f k2} (x+2)^{\f{l+m}2 + \f k2}. \label{SerTilS}
\ee

It is easy to see that the lowest orbital momentum, will give the main
contribution to the energy of radiation. The zero orbital momentum, $l=0$,
does not give contribution to the spectral density \Ref{SpeDen} because in
this case $m=0$. The main contribution comes from the term $l=1,\ (m=\pm 1)$.
Taking into account only this orbital momentum we arrive at the formula
\be
\f{d {\cal E}}{d\omega} = (1-\alpha)^2\f{\pi q^2 v^2}{54} e^{-2 \rho \omega/v}
\left(49 + (1+2\f{\rho \omega}v)^2\right). \label{dE/dwNon}
\ee

It is worth to note that the main contribution to energy comes
from the following interval of frequency
\be
0\ \leq \omega \leq     \f v{2\rho},
\ee
which is due to the exponent in Eq. \Ref{SerTilS}. The maximum of
energy is at zero frequency. Integrating Eq. \Ref{dE/dwNon} we
recover the formula for total energy in the non-relativistic case
\Ref{ENon}.

For the ultra-relativistic case, the expression for the spectral
density is more complicate. We may estimate the interval of frequency from
the analysis of the integrand for $\tilde{S}^l_m$:
\be
\tilde{S}^l_m =\int_1^\infty dy e^{-\Omega y}\left(\f{y+1}{y-1}\right)^{-\f
m2}\f{I_{l+1/2}(\Omega v \sqrt{y^2 -1})}{(y^2 - 1)^{1/4}}.  \label{TilSUlt}
\ee

In this case $v\to 1$ and as a consequence all interval of
integration in Eq. \Ref{TilSUlt} is important. Because the Bessel
function tends exponentially to infinity for large argument, one
has the following exponent in the integrand
\bd
e^{-\Omega (y - v\sqrt{y^2 -1})},
\ed
whose maximum contribution comes from the minimum of the function
$ y - v\sqrt{y^2 -1}$. The minimum of this function is at the
point $y_*=\gamma$ and the main contribution from the exponent due to
this point is
\bd
e^{-\Omega/\gamma}.
\ed
Taking into account this estimation we observe that in the ultra-relativistic
case the domain of frequencies is
\be
0 \leq \omega \leq \f{\gamma}{2\rho}.
\ee

It is worthy noticing that the  intervals of frequencies in both cases,
non-relativistic and ultra-relativist, are exactly the same
that corresponding to the cosmic string space-time.

%%%%%%%%%%%%%%%%%%%%%%%%%%%%%%%%%%%%%%%%%%%%%%%%%%%%%%%%%%%%%%%%%%
\section{Conclusion}\label{Sec3}
%%%%%%%%%%%%%%%%%%%%%%%%%%%%%%%%%%%%%%%%%%%%%%%%%%%%%%%%%%%%%%%%%%
We investigated the radiation emitted by a scalar particle moving
along a geodesic line in the point-like global monopole
space-time. This emission of radiation  arises due to the
geometric and to the topological features of this space-time.
Considering the case of a scalar field minimally coupled with
gravity and a specific situation in which the solid angle deficit
is small we find that the total energy radiated by a particle
along its trajectory is proportional to the cube of the velocity
and to the cube of the Lorenz parameter in the non-relativistic
and ultra- relativistic cases, respectively

The spectral density of radiation in the non-relativistic case has a maximum
at zero frequency with upper bound $v/2\rho$ and for the ultra-relativistic
case, the upper bound is $\gamma /2\rho$.

Let us now compare the energy of the radiation emitted by a uniformly
moving charged scalar particle with the one due to the self-force. The
self-energy, $U_s$, of a charged scalar particle at the rest in a
global point-like monopole space-time is given by

\bd
U_s = \f{q^2}{4r} S(\alpha),
\ed
which was obtained dividing by two the corresponding result
in \cite{BezFur97}, due to the fact that in the present case we are
considering a charged scalar particle.

For small solid angle deficit, the function $S(\alpha) \approx
\f\pi 8 (1-\alpha)$ and therefore
\bd
U_s \approx \f{\pi q^2}{32r}(1-\alpha).
\ed

This potential has the same form as the Coulomb interaction
(repulsion) between two particles with charges $e_1=q$ and $e_2=\pi q
(1-\alpha)/32$. Because the potential is proportional to the small
parameter $(1-\alpha)$ we can consider, approximately, the particle as
moving in the flat Minkowsky space-time and we may use standard results for
energy radiation of the Coulomb scattering problem.

For an ultra-relativistic particle (see \S 73 in
Ref.\cite{Landau}) we have the following expression for the total
emitted energy
\be
{\cal E}_s^{ultra} = \f{\pi^3}{2048} \f{q^6 (1-\alpha)^2
\gamma^2}{m^2\rho^3}. \label{EsUltra}
\ee
We would like to call attention to the fact that this energy depends on the
mass of the particle and this is connected with the fact that the trajectory
of particle is not a geodesic.

Let us now compare results in \Ref{EsUltra} with \Ref{EUltra}, as follows
\bd
\f{{\cal E}_s^{ultra}}{{\cal E}^{ultra}} = \f 1{192} \f 1{\rho_c^2
\gamma},
\ed
where $\rho_c = \rho /\lambda_c$ is the impact parameter $\rho$
measured in Compton wavelengths of the scalar particle $\lambda_c
= q^2/m$. The classical electrodynamics which we are using
is valid for distances much greater than the Compton wavelength, that is,
for $\rho_c \gg 1$. Therefore, due to this fact and because $\gamma
\gg 1 $, we have that
\bd
{\cal E}_s^{ultra} \ll {\cal E}^{ultra}.
\ed
which means that the bremstrahlung of an ultra-relativistic particle prevails
over the radiation due to the self-interaction.

In the non-relativistic case (see \S 70 in Ref.\cite{Landau}) we have the
following ratio

\be
\f{{\cal E}_s^{non}}{{\cal E}^{non}} = \f{f(p)}{3\pi (1 -
\alpha)^2},\label{EsNon}
\ee
where
\bnn
f(p)&=&p^2 \left[(\pi - 2 \arctan p) (1+3p^2) - 6p\right],\\
p&=&\f{1-\alpha}{v^2\rho_c}.
\enn
The function $f(p)$ tends to zero as $\pi p^2$ and to infinity as $
8/15p$. It has maximum at $p\approx 1$ with value $f(1) \approx 0.3$.

The ratio \Ref{EsNon} may be both greater and smaller then unit.
Formal limits of small velocities $v\to 0\ (p\to \infty)$ or great
value of impact parameters $\rho_c\to \infty\ (p\to 0)$ show that
the ratio \Ref{EsNon} tends to zero and in these regions the
bremsstrahlung prevails over the radiation due to self-interaction.
Nevertheless there are some ranges of the velocities and the impact
parameter in which the radiation due to the self-interaction prevails over
that coming from the bremsstrahlung. For $(1-\alpha) \ll
v^2 \ll 1 $, we have $p\ll 1$ for arbitrary $\rho_c >1$ and thus

\bd
\f{{\cal E}_s^{non}}{{\cal E}^{non}} \approx \f 1{3 v^4 \rho_c^2}.
\ed
Therefore, this ratio is greater than unit if the impact parameter
assumes values in the interval
\bd
1\ll \rho_c \ll \f 1{\sqrt{3} v^2}.
\ed

As a conclusion we can say that particles moving along geodesic
lines in the space-time of a point-like global monopole will emit
radiation in the same way as in case of an infinitely thin cosmic
string space-time \cite{AliGal89}. As in the case of an infinitely
thin cosmic string space-time, the energy emitted depends on the
angle deficit and vanishes when this angle deficit vanishes, but
in the present case, this radiation arises associated with the
curvature and non-trivial topology of the space-time of the global
monopole, differently from the cosmic string case in which the
effect comes exclusively from the non-trivial topology of the
space-time.

We have focused our attention not on astrophysical implications of the
obtained results but on the features of the global monopole on the
radiation phenomena. Nevertheless, the application of these results
may have significance from the astrophysical point of view because this
radiation mechanism could give rise large energy releases under some
conditions relevant in an astrophysical scenario.

Finally, it is interesting to call attention to the fact that
based on our results we can obtain the ones corresponding to an electrically
charged particle if we just multiply these by two in order to take into
account the different polarizations of photons.

%%%%%%%%%%%%%%%%%%%%%%%%%%%%%%%%%%%%%%%%%%%%%%%%%%%%%%%%%%%%%%%%%%%%%
\section*{Acknowledgment}
NK is grateful to Departamento de F\'{\i}sica, Universidade
Federal da Para\'{\i}ba (Brazil), where this work was done, for
hospitality. This work was supported in part by CAPES and by BFFR
No. 02-02017177. VBB also would like to thank Conselho Nacional de
Desenvolvimento Cientifico e Tecnol\'ogico (CNPq) for partial
financial support.

%%%%%%%%%%%%%%%%%%%%%%%%%%%%%%%%%%%%%%%%%%%%%%%%%%%%%%%%%%%%%%%%%%%%%

%%%%%%%%%%%%%%%%%%%%%%%%%%%%%%%%%%%%%%%%%%%%%%%%%%%%%%%%%%%%%%%%%%%%%

\end{document}